\documentclass[11pt]{article}
\usepackage{geometry}
\usepackage{cite}
\usepackage{amsmath}
\usepackage{amssymb}
\usepackage{amsfonts}
\usepackage{graphicx}
\usepackage{caption}
\usepackage{subcaption}
\usepackage{float}
\usepackage[usenames,dvipsnames]{xcolor}
\usepackage{tikz}
\usepackage{pstricks}
\usepackage{authblk}
\usepackage{verbatim}
\usepackage{setspace}
\usepackage{gensymb}
\begin{document}
\date{}
\author{Pritam Banerjee \thanks{bpritam@iitk.ac.in}}
\author{Debojyoti Garain \thanks{dgarain@iitk.ac.in}}
\author{Suvankar Paul \thanks{svnkr@iitk.ac.in}}
\author{Rajibul Shaikh \thanks{rshaikh@iitk.ac.in}}
\author{Tapobrata Sarkar \thanks{tapo@iitk.ac.in}}
\affil{Department of Physics, \\ Indian Institute of Technology, \\ Kanpur 208016, India}
\title{Constraining modified gravity from tidal phenomena in binary stars}
\maketitle
\begin{abstract}
In beyond-Horndeski theories of gravity, the Vainshtein screening mechanism might only be partially effective inside stellar 
objects. This results in a modification of the pressure balance equation inside
stars, often characterized by a single parameter ($\Upsilon$) in isotropic systems. We show how to constrain such 
theories of modified gravity, using tidal effects. We study such effects in cataclysmic variable star binaries and 
numerically obtain limits on the critical masses of the donor stars, below which they are tidally disrupted, 
by modeling them in beyond-Horndeski theories. This is contrasted with values of the donor masses, obtained 
using existing observational data, by a Monte Carlo error progression method.
A best fit scenario of the two yields a parametric constraint in the theories that we consider, within the approximations used. 
Here, we obtain the allowed range $ 0 \le \Upsilon \le 0.47 $. 

\end{abstract}

\section{Introduction}
\label{sec-1}
Einstein's theory of general relativity (GR) remains one of the most profound and successful theories of gravity till date, 
and has withstood various precision tests over the last century. Yet, of late, issues related to the observed cosmic
acceleration and the cosmological constant have been shown to indicate that modifications to GR are possibly necessary. 
Such ``modified gravity'' scenarios are becoming increasingly popular in the literature (see, e.g., the reviews
in \cite{CliftonRev}, \cite{LangloisRev}, \cite{IshakRev}, \cite{KaseRev}), and have become 
robust tools for a theoretical understanding of dark matter and dark energy. Naturally, modified gravity scenarios
are constrained via observations, and several such constraints have been proposed and discussed in the recent literature,
with some of the most important ones arising out of the detection of gravitational waves GW170817 and GRB170817A.
The purpose of this paper is to provide a novel methodology wherein one can constrain a class of modified gravity 
scenarios called beyond-Horndeski theories (discussed below, for a recent review, see \cite{KobayashiRev}) from 
tidal effects in cataclysmic variable (CV) binary stars.  

An important artifact of beyond-Horndeski theories of gravity is that the Vainshtein mechanism is only partially
effective inside a stellar object. Remarkably, this means that for stars without pressure asymmetry, the internal pressure
balance equation is modified from its usual form, with the departure from the Newtonian limit of GR being 
characterized by a single constant in most cases. Astrophysical scenarios might thus be suitable in constraining this parameter
by comparing with experimental data, since most stellar observables use the pressure balance equation
in their derivations. Indeed, a number of important works have appeared in the literature
in the last few years, which attempt to find bounds on the modified gravity parameter via different astrophysical
scenarios, in low mass brown or white dwarf stars. Such bounds then translate into ones for viable theories
of modified gravity. 

In this paper, we propose a novel scenario for constraining modified gravity through tidal effects in CV stars. 
Broadly, these are binary systems, with a primary white dwarf that can accrete matter from a Roche lobe filling 
secondary donor star. We model the donor star in beyond-Hordneski theories via  a modified interior pressure balance
equation, as discussed above.  
Since outside the donor star the Vainshtein mechanism works perfectly, we use the tidal 
potential by approximating the spacetime in which the donor moves as
a Schwarzschild background, neglecting any back-reaction (the effects of rotation of the
primary can be neglected to a first approximation). As appropriate in GR, the tidal effects on the secondary donor star are
computed by constructing a tetrad called the Fermi normal frame which, by definition, gives locally flat spacetime
along the geodesic trajectory of the donor. By a numerical analysis specialized to such a locally flat spacetime, 
we then compute the critical value 
of the central density of the donor, so that it fills its Roche lobe. The critical density profile is then used to calculate the value
of the critical mass of the donor, below which it is tidally disrupted. 
This is done for $13$ different CV systems for which we have taken the data from 
existing literature (details of these are provided in an appendix to this paper, for convenience of the reader). In
these systems, the orbital time periods of rotation of the donor stars around their primaries are less than $6$ hours, 
for which it is known that one can use a polytropic equation of state with a polytropic index $3/2$, to a good approximation. 
Using this value of the polytropic index, we find how the values of the critical masses of the donor stars
depend non-trivially on the modified gravity parameter. Next, using the Monte Carlo error progression method, 
we obtain the mass ranges for the secondary donor stars from observational data. By comparing
the critical values of the masses of the donors obtained via the numerical procedure, with the observed secondary masses,
and using a chi-square best fit scenario, we are able to obtain bounds on the modified gravity parameter. 
This is our broad methodology which will be elaborated in the rest of the paper. 

This paper is organized as follows. In the next section (\ref{prelim}), we discuss the mathematical 
preliminaries and the setup of the problem. In section (\ref{numerical}), we will elaborate upon the numerical
procedure to study tidal disruption of a polytropic star in modified gravity, and also establish the 
dependence of the critical mass of a donor star on the modified gravity parameter. Section (\ref{obs}) deals with the
observational data on the CV stars and here we elaborate upon the Monte Carlo procedure for obtaining
the mass of the secondary, which is compared with that obtained from our numerical analysis 
in section (\ref{comparison}). We end the paper with our conclusions and some discussions 
in section (\ref{conclusions}), followed by an appendix with relevant details of the CV binaries considered here. 

\section{Preliminaries and setup}\label{prelim}

In four dimensions, Lovelock's theorem
guarantees that the Einstein's equations are the only possible second order equations that can be derived
from a Lagrangian density that depends only on the metric and its first and second derivatives, and is linear in the
second derivatives. Hence, the minimal modification to gravity 
involves adding an extra degree of freedom, the simplest being a single scalar called the Galileon. As the name
suggests, this has a shift symmetry in flat backgrounds, and possible Lagrangians can
be constructed so that the equations of motion are of second order. In four dimensions, the most general 
Lagrangian that respects these conditions has been written down in \cite{Nicolis}, and its covariant version
which breaks the shift symmetry while retaining second order equations of motion is constructed
in \cite{Deffayet}. A further generalization of the theory (called the generalized Galileon) 
appeared in \cite{Deffayet1} and was shown in \cite{Kobayashi1} to be equivalent to Horndeski's
Lagrangian \cite{Horndeski}, written in the 70's. 

Physicality of extensions of GR is greatly limited by the fact that there should be no propagating ghost 
degrees of freedom. This is intimately related to the Ostrogradsky theorem, which broadly states that non-degenerate
Lagrangians that depend on second and higher derivatives of the scalar field have ghost instabilities. This is
in fact the reason that one usually restricts to extensions of GR that involve up to second order equations 
of motion. Importantly however, if the Lagrangian is degenerate, then one can bypass the Ostrogradsky theorem 
and reduce possible higher derivative equations of motion to second order ones. These are the scalar tensor
theories beyond Horndeski. Although several such theories have been proposed, they are tightly constrained
by recent detection of gravitational waves GW170817 (and also GRB170817A). 

Indeed, the pioneering works of \cite{BH1}, \cite{BH2}, have shown how 
to extend the generalized Galileon theories to the beyond-Horndeski class of theories, that also have no 
ghost modes in their spectra. These models are called the GLPV models, or the $G^3$ Galieon. 
This has been further generalized in the important works of \cite{LangloisNoui}, \cite{LangloisNoui1} (see also \cite{DHOST1})
to models which are free from Ostrogradsky instabilities, and in which the GLPV models arise as special cases. 
The key issue here is that the ghost instability can be
avoided if the Lagrangian is degenerate (i.e its Jacobian matrix constructed out of the higher derivative terms vanishes). 
In such a situation, the higher order equations of motion can be reduced to a system of second order equations. 
These theories are dubbed as the degenerate higher order scalar tensor (DHOST) theories beyond Horndeski
and have been extremely well studied in the recent past.  

It is clear that, in such general situations, the compatibility of modifications to GR effects 
in solar system tests necessitate invoking some kind of screening mechanism. As GR is tested extremely well
at such small scales, such mechanisms are necessary to screen the effects of modified gravity at such scales.
While several possibilities exist that hide the effect of modified gravity at small scales, one
of the most efficient is the Vainshtein mechanism \cite{Vainshtein} (see, e.g. \cite{JainKhoury}, 
\cite{BabichevRev} for reviews), where GR is recovered in such near regimes, via a non-linear screening of modified gravity. 
The breaking of the Vainshtein mechanism is central to our discussion on modified gravity, and the work 
of \cite{KWY} have shown that this can be non-trivial inside a source. That is, inside a stellar object,
Vainshtein screening is only partially effective. In the wake of GW170817, this issue was revisited in
\cite{VainGW1} and \cite{VainGW2}. We follow the notations of \cite{VainGW1}. 
Incorporating previous results of \cite{LangloisNoui1}, one writes down
the general Lagrangian
\begin{equation}
{\mathcal L} = G\left(\phi,X\right)R + \sum_{I=1}^5 A_I{\mathcal L}_I~,
\end{equation}
where $R$ is the four dimensional Ricci scalar, and one introduces a non-minimal coupling with gravity in the first term,
with $G$ being a function of $\phi$ and $X = \phi^{\mu}\phi_{\mu}$. In GR, $G$ is the Newton's constant, which
is renormalized here. 
The remaining terms take the form
\begin{eqnarray}
{\mathcal L}_1 = \phi_{\mu\nu}\phi^{\mu\nu}~,~~{\mathcal L}_2 = \left(\Box\phi\right)^2~,~~
{\mathcal L}_3 = \phi^{\mu}\phi_{\mu\nu}\phi^{\nu}\Box\phi~,~~
{\mathcal L}_4 = \phi_{\mu\rho}\phi^{\rho\nu}\phi^{\mu}\phi_{\nu}~,~~
{\mathcal L}_5 = \left(\phi^{\mu}\phi_{\mu\nu}\phi^{\nu}\right)^2~,
\end{eqnarray}
and the coefficients $A_I$ are general functions of $\phi$, with $\phi_{\mu}=D_{\mu}\phi$,
$\phi_{\mu\nu}=D_{\mu}D_{\nu}\phi$, and $\Box\phi=D^{\mu}D_{\mu}\phi$, where $D_{\mu}$ denotes
the covariant derivative.
The coefficient $A_1=0$ as follows from the GW170817 constraint that the speed of the gravitational
wave in this theory equals the speed of light (with an error of one part in $10^{15}$). Moreover, to
avoid Ostrogradsky instabilities, one has to set $A_2=0$ as well, and further $A_4$ and $A_5$ get related
to $A_3$ via $X, G$ and $\partial G/\partial X$. One then has $G$ and $A_3$ as free parameters. 
Now we consider a generic Friedman-Robertson-Walker metric in the flat space-time limit, given in the spherically
symmetric case by
\begin{equation}
ds^2 = -\left(1 + 2\Phi(t,r)\right)c^2dt^2 + a(t)^2\left(1-2\Psi(t,r)\right)\left[dr^2 + r^2\left(d\theta^2 + \sin^2\theta d\phi^2\right)\right]~,
\label{lineelement}
\end{equation}
with a scale factor $a(t)$, and consider perturbations of the scalar field about this background. Assuming spherical symmetry, in the 
limit that the radial coordinate is much less than the Vainshtein radius, one obtains after some algebra \cite{VainGW1},\cite{VainGW2}
\begin{equation}
\frac{d\Phi}{dr} = \frac{G M_r}{c^2r^2} + \frac{\Upsilon_1}{4}\frac{G}{c^2}\frac{d^2M_r}{dr^2}~,~~
\frac{d\Psi}{dr} = \frac{G M_r}{c^2r^2} - \frac{5\Upsilon_2}{4}\frac{G}{c^2 r}\frac{d M_r}{d r}
+ \Upsilon_3\frac{G}{c^2}\frac{d^2M_r}{dr^2}~,
\label{phipsider}
\end{equation}
where a quasi-static approximation is used, i.e., the time derivatives of the perturbations are considered to
be small compared to the spatial ones, and $M_r$ denotes the mass of the stellar object up to radius $r$. 
Here, $\Upsilon_i$, $i=1,\cdots 3$ are specific functions of 
$A_3$, $X, G$ and $(\partial G/\partial X)$, and so is the renormalized Newton's constant. 
This general scenario reduces to the result of \cite{KWY} when $A_3X=-4(\partial G/\partial X)$ in which case 
$\Upsilon_3=0$. 

A remarkable consequence of the above is that, in the low energy (Newtonian) limit, the pressure balance equation 
inside astrophysical objects is modified (outside such objects, with $dM/dr=d^2M/dr^2 = 0$, usual GR is recovered). 
Since the pressure balance equation is a crucial ingredient 
in the derivation of analytical formulas corresponding to observables that can be experimentally verified, 
one is immediately led to the conclusion that the theoretical 
results of this class of modified gravity theories can be constrained by experimental data. 
To keep the discussion general at this stage, we will start by taking the Newtonian limit of the stress tensor inside
a stellar object, that in GR is given by $T^{\mu}_{\nu} = {\rm diag}(-\rho c^2, P_{rad}, P_{\perp}, P_{\perp})$ where
$c$ is the speed of light and we have allowed for the fact that, in general, an anisotropy might be allowed, with
$P_{rad}$ being the radial pressure and $P_{\perp}$ being the tangential one, and spherical symmetry dictating that
such tangential pressures along the non-radial directions should be equal.

Now, we will use Eq. (\ref{lineelement}) with the scale factor set to unity, since we are interested only in a static situation. Here, 
$\Phi(r)$ is the Newtonian potential, and we will take $\Phi(r),\Psi(r) \ll 1$. That the energy momentum tensor is
covariantly conserved, i.e., $D_{\mu}T^{\mu\nu}=0$ (with $D_{\mu}$ being the covariant derivative), then gives
in the Newtonian limit, 
\begin{equation}
\frac{dP_{rad}}{dr} = -\rho c^2\frac{d\Phi}{dr} +\frac{2}{r}\left(P_{\perp}-P_{rad}\right)\left(1-r\frac{d\Psi}{dr}\right)=0~.
\label{TOVA2}
\end{equation}
Terms involving $\Psi$ can only come into the picture for theories with an anisotropy. 
With the differential equations for the potentials $\Phi(r)$ and $\Psi(r)$ now given by Eq. (\ref{phipsider}),
substituting Eq. (\ref{phipsider}) in Eq. (\ref{TOVA2}), and assuming an isotropic situation 
$P_{rad} = P_{\perp}=P$, we obtain the final form of the pressure balance equation
\begin{equation}
\frac{dP}{dr} = - \frac{GM_r\rho}{r^2} - \frac{\Upsilon}{4}G\rho\frac{d^2M_r}{dr^2}~,~
\label{hydrostatic}
\end{equation}
where the mass conservation equation defines the density, i.e.,
\begin{equation}
\frac{dM_{r}}{dr}=4\pi r^{2}\rho ~.
\label{mass_conservation_equn}
\end{equation}
Here $G$ is the Newton's constant and as before, $M_r$ denotes the mass of a stellar object up to radius $r$, and 
the second relation gives the mass in terms of the density.  Also, for ease of notation, we have 
called $\Upsilon_1 = \Upsilon$. In the simplest possible scenario, when $\Upsilon$ is a constant, one
can put astrophysical constraints on it from stellar scenarios. 

Indeed, a plethora of  activities have been reported in the last few years on these lines. While the pioneering work of Koyama and
Sakstein \cite{SaksteinPRD} reported results on the luminosity and temperature of main sequence stars in 
modified gravity and contrasted them with their 
GR cousins, \cite{Saito} performed a generic analysis of the Lane-Emden equation inside astrophysical objects
in the context of modified gravity and came up with constraint $ \Upsilon > -2/3 $. 
Further, \cite{Sakstein}, \cite{Sakstein3} put a bound on this parameter by studying the minimum mass of Hydrogen
burning in brown dwarf stars, and predicted $ \Upsilon \leqslant 1.6 $. 
White dwarfs were comprehensively analyzed in \cite{Jain} which refined these
bounds as $ -0.48 \leqslant \Upsilon \leqslant 0.54 $ with $ 5\sigma $ and 
$ -0.18 \leqslant \Upsilon \leqslant 0.27 $ with $ 1\sigma $ confidence level. 
A stronger constraint, i.e. $ \Upsilon \leqslant 0.18 $, is given by \cite{Saltas} from white dwarf scenarios.
In \cite{Babichev}, the lower bound $\Upsilon > -0.44$ was established via considerations of strong gravity, 
and \cite{Tapo1} reported a lower bound of $\Upsilon \geqslant -0.12$ from analysis of the physics of brown
dwarfs. For a recent review of the effects of modification of gravity in stellar objects, see \cite{GonzaloRev}.

\section{Tidal deformation of a polytropic star in modified gravity}
\label{numerical}

We will begin with the pressure balance equation of Eq. (\ref{hydrostatic}) with the definition of the density
given in Eq. (\ref{mass_conservation_equn}) inside a stellar object. These equations are used to find the 
density profile inside an astrophysical source. In order to do this, 
we first consider a polytropic equation of state, $ P= K\rho^{1+\frac{1}{n}} $ where $ P $ is the pressure and 
$ \rho $ is the density. $ n $ is the polytropic index and $ K $ is known as the polytropic constant. We also 
define a dimensionless quantity $ \xi $ as $ r=\bar{r}\xi $ where $ \bar{r} $ is a constant. We now set the 
density $ \rho $ as a function of $ \xi $ by the relation $ \rho=\rho_c\theta(\xi)^n $. The central density is 
denoted by $ \rho_c $. These relations are put in Eqs. (\ref{hydrostatic}) and (\ref{mass_conservation_equn}) 
to obtain the modified Lane-Emden equation   
\begin{equation}
\frac{1}{\xi^2}\frac{d}{d\xi}\left[\left(1+\frac{\Upsilon}{4}n\xi^2\theta^{n-1}\right)\xi^2\frac{d\theta}{d\xi}
+  \frac{\Upsilon}{2}\xi^3\theta^n\right]=-\theta^n~.
\label{modified_lane_emden_equn}
\end{equation} 
The constant $ \bar{r} $ is obtained as 
\begin{equation}
\bar{r}=\left[\frac{K\rho_c^{\frac{1}{n}-1}(n+1)}{4\pi G}\right]^{\frac{1}{2}}~.
\end{equation}
When $ \Upsilon = 0 $, we get back the Lane-Emden equation for GR. The density profile is obtained as a solution of 
Eq. (\ref{modified_lane_emden_equn}) for a fixed value of the parameter $ n $. In order to solve the equation, 
the boundary conditions are fixed as $ \theta(0)=1 $ and $ \theta'(0)=0 $. These conditions suggest that, for a 
spherically symmetric static astrophysical object, the central density is $ \rho_c $ which is the maximum value 
in the entire density profile. 
  
We consider a star which is described by the modified Lane-Emden equation. Its density profile depends 
on the polytropic index and the value of the modified parameter $ \Upsilon $. It suggests that a star having 
a fixed radius can have different masses depending on the different values of $ \Upsilon $. This is evident from 
the expression 
\begin{equation}
M = \int_{0}^{R}\rho(r) 4\pi r^2 dr  = \bar{r}^3\rho_c\int_{0}^{\xi_R} \theta(\xi)^n 4\pi \xi^2 d\xi ~.
\end{equation} 
Here, $ R $ is the radius of the star and $ \xi_R $ is the value of $ \xi $ at the surface. 
According to Eq. (\ref{modified_lane_emden_equn}) the density profile $ \theta(\xi) $ varies, depending on $ \Upsilon $. 
As a result, $ M $ depends on $ \Upsilon $.
   
Now, when a star experiences a tidal force field, it undergoes tidal deformation. It can even be tidally 
disrupted when the tidal field is strong enough to overcome the self gravity of the star. The limit at which the 
tidal force at the surface of the star is equal to its self gravity, is called the tidal disruption limit or the 
Roche limit. At this limit, the star fills its Roche lobe and remains just stable without being tidally disrupted. If a star of 
mass $ M $ orbits a black hole or any other compact object of mass $ M_{BH} $ whose exterior can be approximated
by a black hole space-time, it can be tidally disrupted if its 
orbital distance is less than the tidal radius $ r_t $. A phenomenological value of $ r_t $ was first given in the 
work of \cite{Kopal}, whereas \cite{Pacynski} provided a relation for $r_t$ in terms of the mass ratio $ q=M/M_{BH} $
by studying the Roche geometry. The work of Eggleton \cite{Eggleton} improved the relation to give 
\begin{equation}
\frac{R_v}{r_t} = \frac{0.49 q^{2/3}}{0.6q^{2/3}+\ln\left(1+q^{1/3}\right)}~,
\label{EForm}
\end{equation}
where $ R_v $ is the volume equivalent radius. This formula will be useful for our discussion to follow, and
we will call it the Eggleton formula (EF). 
Of course the EF does not include the effects of modified gravity. In the following, we will discuss a
methodology for finding the tidal disruption limit in the context of modified gravity. This method will be used later 
on, to put a constraint on the values of $ \Upsilon $. 

We consider a star described by the modified Lane-Emden equation, is deformed under the influence 
of a tidal field. We want to find the maximum deformation that the star can attain without getting tidally disrupted. 
In other words, we find the minimum possible value of the central density $ \rho_c $ of the star. We start by 
considering a polytropic fluid star in a Fermi normal frame \cite{Manasse-Misner}. In the absence of tidal field, the density 
profile of the star is spherically symmetric as obtained from Eq. (\ref{modified_lane_emden_equn}). The tidal 
potential up to a fourth order approximation is expressed in the Fermi normal frame as \cite{Ishii-Kerr} (the dummy index $i$ used here is not to be 
confused with the inclination angle of the next section)
\begin{equation}    
\phi_{\text{tidal}} =\frac{1}{2} C_{ij} x^i x^j + \frac{1}{6} C_{ijk} x^i x^j x^k + \frac{1}{24} \left[ C_{ijkl} + 
4 C_{\left(ij\right.} C_{\left.kl\right)} - 4 B_{\left(kl |n|\right.} B_{\left.ij\right)n} \right] x^i x^j x^k x^l + O(x^5)~,
\label{eq.phi_tidal}    
\end{equation}
where,  \{$x^0,x^1,x^2,x^3$\} are the Fermi normal coordinates. The coefficients are obtained from the Riemann tensor as
\begin{equation}
C_{ij} = R_{0i0j}, ~~~ C_{ijk} = R_{0\left(i|0|j;k\right)}, ~~~ C_{ijkl} = R_{0\left(i|0|j;kl\right)}, 
~~~ B_{ijk} = R_{k\left(ij\right)0}~.
\label{eq.Cij}
\end{equation}
In these expressions above, $ i,j,k,\cdots $ are spatial indices. Partial and covariant derivatives are indicated 
by the symbols  ` , '  and  ` ; '  respectively. $  R_{0\left(i|m|j;kl\right)} $ denotes a summation over all the possible 
permutations of the indices $i,j,k,l$ with $m$ fixed at its position, divided by the total number of permutations.   
In the presence of a tidal force field, the deformation is obtained by numerically solving the Euler equation in the 
Fermi normal frame, given by
\begin{equation}
\rho \frac{\partial v_i}{\partial \tau} + \rho v^j \frac{\partial v_i}{\partial x^j} = - \frac{\partial P}{\partial x^i} - 
\rho \frac{\partial (\Phi + \phi_{\text{tidal}})}{\partial x^i} +\rho \left[ v^j \left( \frac{\partial A_j}{\partial x^i} - 
\frac{\partial A_i}{\partial x^j} \right) - \frac{\partial A_i}{\partial \tau} \right]~,
\label{eq.hydrodynamic}
\end{equation}  
where $ v^i $ is the velocity field of the star, which is rotating in the Fermi normal frame. 
The part containing the vector potential $A_k = \frac{2}{3} B_{ijk} x^i x^j$ corresponds to the  gravito-magnetic 
force field. The self gravitational potential $ \Phi $ satisfies the modified Poisson equation which is obtained 
using Eqs. (\ref{TOVA2}) to (\ref{mass_conservation_equn}) as (with $c=1$),
\begin{equation}
\nabla^2 \Phi = 4 \pi G \rho + \Upsilon G\left(6\pi \rho + 6 \pi r \frac{d\rho}{dr} + \pi r^2 \frac{d^2\rho}{dr^2}\right)~.
\label{eq.poisson}
\end{equation}
It is to be noted that the modified part contains the density as well as its first and second derivatives. 
It reduces to the usual form when $ \Upsilon = 0 $. We solve Eqs. (\ref{eq.hydrodynamic}) and (\ref{eq.poisson}) 
to get the deformed shape of the fluid star in equilibrium in the presence of the tidal field. Firstly, the spherically 
symmetric density profile and its derivatives, as obtained from Eq. (\ref{modified_lane_emden_equn}), is given 
as input to find $ \Phi $ from Eq. (\ref{eq.poisson}). Then, $ \Phi $ is used in Eq. (\ref{eq.hydrodynamic}) to obtain 
the updated density profile $ \rho $. This procedure is repeated until the desired convergence is obtained. A full 
numerical description is given in \cite{BANERJEE201929}. We find the minimum possible central density $ \rho_{\text{crit}} $
for which the star just remains stable in the tidal field. If the central density is below
$ \rho_{\text{crit}} $, the star is tidally disintegrated. In other words, the critical density profile is calculated 
at the tidal disruption limit or Roche limit.  By integrating the critical density profile we obtain the critical mass 
$ M_{\text{crit}} $, such that donor stars having mass less than 
$ M_{\text{crit}} $ are tidally disrupted. The tidal disruption limit is thus obtained by the following input parameters : (a) the mass of
the central object, (b) the orbital distance of the star, and (c) the configuration of the star. Next, we discuss how the 
critical mass $ M_{\text{crit}} $ depends on the modified parameter $ \Upsilon $. 

We consider a star moving around a Schwarzschild black hole in a circular orbit. The choice of the orbit 
fixes the energy and angular momentum of the star, approximated as a test particle. These are used to obtain the Fermi normal 
basis which can be defined throughout the entire geodesic of the star. The 
$4$-velocity of the star is the timelike basis vector of the Fermi normal frame. This suggests that the frame 
moves with the star along its timelike geodesic. Hence, it is convenient to solve the fluid equations in the 
Fermi normal frame to find the tidal deformation of the star.
\begin{figure}[ht]
\centering
\includegraphics[scale=0.6]{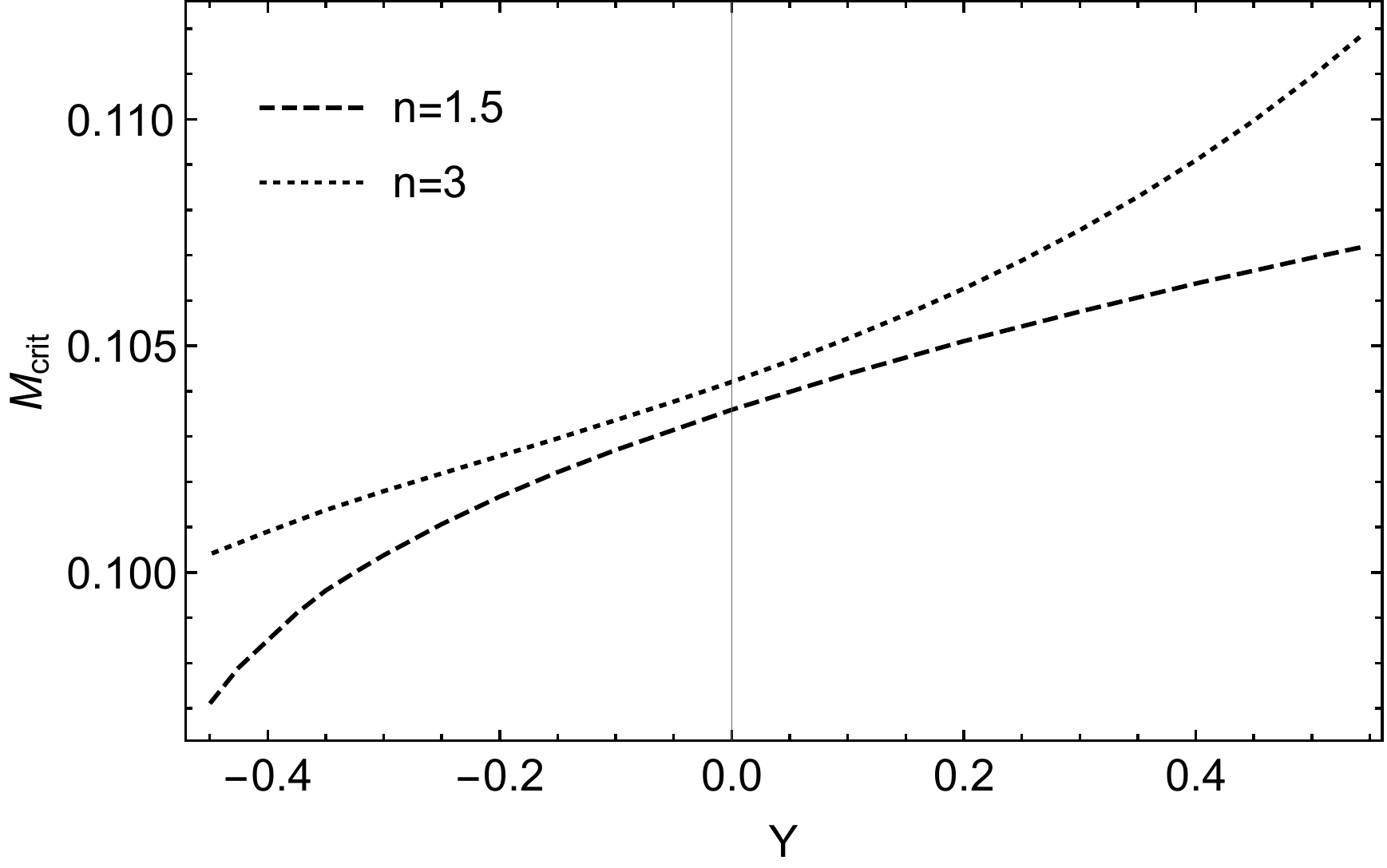}
\caption{Critical mass $ M_{\text{crit}} $ increases with $ \Upsilon $ for a fixed orbital radius $ a = 5\times 10^3 M_{BH} $ and 
$ R = 10^3 M_{BH}$. Here, units of $ c=G=M_{BH}=1 $ are assumed.} 
\label{fig1}
\end{figure}
	
Fig. \ref{fig1} shows the variation of $ M_{\text{crit}} $ with $ \Upsilon $. The plot is done at 
orbital radius $ a=5000~M_{BH} $ with $ R=1000~M_{BH} $. The results are shown for two values of the polytropic index,
namely $ n=1.5 $ and $ 3 $. 
It should be mentioned here that the plots are shown in black hole (BH) units, $ c=G=M_{BH}=1 $. 
As Fig. \ref{fig1} indicates, $ M_{\text{crit}} $ increases with $ \Upsilon $. This is consistent with 
the fact that a positive value of $ \Upsilon $ weakens gravity inside an astrophysical body while a 
negative $ \Upsilon $ enhances it. Hence, for a positive $ \Upsilon $, more mass is required to 
keep the star intact from getting tidally disrupted. As a result, the mass of the star at the tidal disruption 
limit increases for positive $ \Upsilon $ and vice versa. 
	
\section{Mass of the secondary star from observational data} 
\label{obs}

We consider a binary system which is described fairly well by the numerical model. The foremost 
requirement is that the secondary star fills its Roche lobe. It is also required to be corotating around the 
primary in a circular orbit. Such assumptions are followed quite well by CV stars 
which consist of a white dwarf primary accreting from a Roche lobe filling secondary star. The secondary 
stars in CVs are likely to be tidally locked to the primary such that the spin velocity of the secondary star is 
similar to its orbital velocity. They are also found to be rotating around the primary in short and fixed periodic 
intervals. Furthermore, the orbital eccentricities of the CVs used in this paper are measured to be 
small \cite{ecentricity}. Hence we consider CV systems as a laboratory where tidal disruption limit can 
be used to test the effect of modified gravity inside astrophysical objects. Our purpose is to find the binary 
parameters such as the masses of the primary and secondary stars ($M_1$ and $M_2$, respectively), the orbital, 
and the volume equivalent radii of the secondary stars ($a$ and $R_2$, respectively) 
by means of observational data. The observed mass range of the secondary is 
then compared with the mass range obtained numerically at the tidal disruption limit by varying the modified parameter. 
As mentioned in the introduction, this procedure allows us to put a constraint on the values of the modified parameter. 
The following discussion relates to that of the previous section with the mass of the primary $M_1 \equiv M_{BH}$ there. 

\begin{table}[h!]
\small
\caption{The list of observed binary parameters used for calculating $M_1$, $M_2$, $a$ and $R_2$ are given. 
Details on the CV systems can be found in the Appendix. \label{table1}}
\begin{tabular}{cccccccccccccc}
\hline
\hline
Name & \multicolumn{8}{c}{~~Known binary parameters~~} \\
& ~~~$ P $~~~ & ~~~~~$ i\degree $~~~~~& ~~~~~$ q $~~~~~ & ~~~~~~~$\Delta\phi_{1/2} $
~~~~~~~ & ~~$ K_1 $~~ & ~~$ K_2 $~~~~& ~$ v\sin i $~& $M_2$ &   \\
& (h) &  &  &  & (km~s$^{-1}$)  &  (km~s$^{-1}$) &  (km~s$^{-1}$) & ($M_\odot$) & \\
\hline
V4140 Sgr & $1.467$ & $80.2\pm0.5$ & $0.125\pm0.015$ & $0.0378\pm0.0005$ & $56\pm7$ & - & - & -  \\
V2051 Oph & $1.5$ & $83.3\pm1.4$ & $0.19\pm0.03$ & $0.0662\pm0.0002$ & $91\pm12$ & - & -  & - \\
OY Car & $1.51$ & $83.3\pm0.2$ & $0.102\pm0.003$ & $0.0506\pm0.0004$ & - & $470\pm2.7$ & -  & - \\
Ex Hya & $1.638$ & $77\pm1$ & - & $0.017\pm0.002$ & $69\pm9$ & $356\pm4$ & - & -  \\
HT Cas & $1.77$ & $81\pm1$& $0.15\pm0.03$ & $0.0493\pm0.0007$ & $58\pm11$ & $389\pm4$  & - & -  \\
IY Uma & $1.77$ & $86\pm1$ & $0.125\pm0.008$ & $0.0637\pm0.0001$ & - & $383\pm6$ & - & -  \\
Z Cha & $1.79$ & $81.78\pm0.13$ & $0.150\pm0.004$ & $0.0534\pm0.0009$ & - & $430\pm16$  & - & - \\
DV Uma & $2.06$ & $84.24\pm0.07$ & $0.151\pm0.001$ & $0.063604$ & - & - & - & $0.15\pm0.02$ & \\
IP Peg & $3.797$ & $81.8\pm0.9$ & $0.45\pm0.04$ & $0.0863$ & - & $298\pm8$ & - & -  \\
UU Aqr & $3.93$ & $78\pm2$ & - & $0.051\pm0.002$ & $121\pm7$ & $327\pm31$ & - & - \\
Gy Cnc & $4.211$ & $77\pm0.9$ & - & $0.060\pm0.005$ & $115\pm7$ & $283\pm17$ & -  &  - \\
Ex Dra & $5.04$ & $85^{+3}_{-2}$ & $0.72\pm0.06$ & $0.1085\pm0.0006$ & - & $210\pm14$ & $140\pm10$ & -  \\
V347 Pup & $5.566$ & $87\pm3$ & - & $0.115\pm0.005$ & - & $198\pm5$ & $130\pm5$ & -  \\
\hline
\end{tabular}
\end{table} 

\begin{table}[h!]
\begin{center}
\small
\caption{$M_1$, $M_2$, $a$ and $R_2$ as obtained from Monte Carlo error progression method 
(using the known binary parameters as shown in Table \ref{table1}) are listed below.  \label{table2}}
\begin{tabular}{cccccccc}
\hline
\hline
~~~~Name~~~~&~~~~~~~~~~~$M_1$~~~~~~~~~~~&~~~~~$a$~~~~~& ~~~~~~~~~~~$R_2$~~~~~~~~~~~&~~~~$M_2$~~~~  \\
& $(M_\odot)$ & $ (R_\odot)$ & $(R_\odot)$ & ~~~~~~~$(M_\odot)$~~~~~~~  \\
\hline
V4140 Sgr & $0.9\pm0.5$ & $0.63\pm0.11$ & $0.13\pm0.02$ & $0.10\pm0.05$   \\
V2051 Oph & $1.2\pm0.9$ & $0.726\pm0.14$ & $0.17\pm0.04$ &$0.22\pm0.11$   \\  
OY Car & $0.834\pm0.015$ & $0.649\pm0.004$ & $0.127\pm0.0017$ & $0.085\pm0.003$ \\	
Ex Hya & $0.49\pm0.03$ & $0.589\pm0.014$ & $0.136\pm0.011$  &$0.095\pm0.017$   \\
HT Cas & $0.62\pm0.04$ & $0.661\pm0.018$ & $0.144\pm0.009$  &$0.09\pm0.02$  \\
IY Uma & $0.55\pm0.03$ & $0.630\pm0.011$ & $0.133\pm0.004$ &$0.068\pm0.006$   \\
Z Cha & $0.84\pm0.09$ & $0.74\pm0.03$ & $0.161\pm0.006$  &$0.125\pm0.014$  \\
DV Uma & $1.00\pm0.13$ & $0.86\pm0.04$ & $0.190\pm0.010$  &$0.15\pm0.02$ &  \\
IP Peg & $0.94\pm0.09$ & $1.37\pm0.05$ & $0.41\pm0.02$ &$0.42\pm0.07$ &  \\
UU Aqr & $1.2\pm0.3$ & $1.48\pm0.11$ & $0.39\pm0.05$ & $0.44\pm0.07$ \\
Gy Cnc & $0.88\pm0.13$ & $1.42\pm0.07$ & $0.41\pm0.03$ &$0.36\pm0.05$  \\
Ex Dra & $0.69\pm0.10$ & $1.58\pm0.08$ & $0.54\pm0.03$ &$0.52\pm0.09$  \\
V347 Pup & $0.63\pm0.08$ & $1.66\pm0.10$ & $0.60\pm0.02$ &$0.53\pm0.13$   \\							
\hline
\end{tabular}
\end{center}
\end{table}

We now specify the set of equations  to find the required binary parameters (see \cite{BT_Mon_1}, \cite{V347_Pup_1}, 
\cite{Monte_Carlo_1} for similar approaches). For the primary mass $ M_1 $, secondary mass $ M_2 $ and the 
orbital period $ P $, the orbital radius $ a $ is given by the Kepler's third law as
\begin{equation}
a = \left[\frac{G(M_1+M_2)}{4\pi^2}\right]^{\frac{1}{3}}P^{\frac{2}{3}}.
\label{MCE_1}
\end{equation}
The mass ratio of the binary $ q $ is defined as 
\begin{equation}
q=\frac{M_2}{M_1} = \frac{K_1}{K_2}~,
\label{MCE_2}
\end{equation}
where $ K_1 $ and $ K_2 $ are the radial velocities of the primary and the secondary stars respectively. 
The volume equivalent radius of the secondary star $ R_2 $ is obtained from primary eclipse light curve. Using geometric 
analysis we have 
(see \cite{R2_formula_1, R2_formula_2, R2_formula_3, R2_formula_4, R2_formula_5, R2_formula_6, R2_formula_7})
\begin{equation}
\left(\frac{R_2}{a}\right)^2 = \sin^2(\pi\Delta\phi_{1/2})+\cos^2(\pi\Delta\phi_{1/2})\cos^2i~,
\label{MCE_3}
\end{equation}
where $ i $ (in degrees) is the inclination of the plane of the binary with respect to the line of sight
(we remind the reader that this should not be confused with the dummy index $i$ used 
in the discussion around Eq.(\ref{eq.phi_tidal})). The difference between
the binary phases at the mid ingress to the mid egress is denoted as $ \Delta\phi_{1/2} $ 
(for details, see, e.g. \cite{DV_Uma_1}). The rotational velocity of the secondary star is obtained 
from the equation  
\begin{equation}
v\sin i = \frac{2\pi\sin i}{P} R_2~,
\label{MCE_4}
\end{equation} 
and is related to $ K_1 $ and $ K_2 $ as 
\begin{equation}
\frac{v\sin i}{R_2} = \frac{K1+K2}{a}~.
\label{MCE_5}
\end{equation}	
We can see that there are six independent equations (Eq.(\ref{MCE_2}) contains two independent
equations) and eleven parameters altogether. Therefore, at least 
five parameters must be known to obtain the others. However, not all the combinations of the five parameters 
are helpful, as can be seen from the above equations. For example, if in a particular combination,
$ a $, $ M_1 $, $ M_2 $ and $P$ are all unknown, we cannot use Keplar's law of eq.(\ref{MCE_1}). 
This reduces the number of independent equations from six to five, which is of little help. Hence we need
to know at least one of these quantities. The binary parameters as obtained from 
observational data are listed in Table \ref{table1}.

For each of the CVs, we need to find $ M_1 $, $ a $, $ M_2 $ and $ R_2 $ by solving Eqs. (\ref{MCE_1}) to 
(\ref{MCE_5}) using the known parameters listed in Table \ref{table1}. We use Monte Carlo error progression 
method with $ 10^5 $ sample values of the known quantities to find the mean and the standard deviation of each of the unknown 
parameters. In order to do that, we consider that the values of the known parameters are 
normally distributed around their measured values and the uncertainties in measurements are set to be the 
standard deviations. For example, in case of V4140 Sgr, the inclination angle $ i $ (in degrees) is considered to follow a normal 
distribution such that its mean is $ 80.2 $ and the standard deviation is $ 0.5 $. Following the known parameter 
distributions, sample values are randomly chosen and used in Eqs. (\ref{MCE_1}) to (\ref{MCE_5}) to find the 
values of the unknown parameters. This is repeated to obtain $ 10^5 $ sets of parameter values, thereby
giving distributions for the unknown parameters. The mean values 
and the standard deviations of the unknown parameters are then calculated using their corresponding distributions. 
Table \ref{table2} shows the values of $ M_1 $, $ M_2 $, $ a $ and $ R_2 $ obtained via Monte 
Carlo error progression using the quantities listed in Table \ref{table1}. Once we know $ M_1 $, $ R_2 $ and $ a $, 
we use these parameters as inputs in the numerical calculation of $M_2^{crit}$ for a fixed value of $\Upsilon$, 
with the condition that the volume equivalent radius of the secondary star at the tidal 
disruption limit is equal to $ R_2 $. Now, by varying $\Upsilon$, and following the same numerical procedure
for each value of $ \Upsilon $, we obtain the critical mass range of the secondary star at the tidal disruption limit 
which is then compared with the observed mass range generated via Monte Carlo. As already mentioned, 
this enables us to put a constraint on the modified parameter.
  	 
We emphasize that the values of $ M_1 $, $ M_2 $, $ a $ and $ R_2 $ for a few binaries used here
are available directly in the literature. However, in these, the authors prefer to use the EF 
(Eq. (\ref{EForm})) which relates the binary parameters at the tidal disruption limit without incorporating the 
modified gravity effects. Since our aim is to put a constraint in the modified parameter, we need to 
use numerical analysis incorporating $\Upsilon $. 
This allows us to find the Roche lobe having different mass values (for a fixed volume equivalent radius) corresponding to 
different values of $\Upsilon$. This set of mass ranges can be used to find a constraint in $ \Upsilon $ only if it is compared 
to another set of masses obtained independent of the tidal phenomena. In this spirit, we do not use the EF. 
We have used the CVs with available eclipse data, such that we get extra information about
the radius of the secondary completely from the light curve and eclipse
width. As a shortcoming of removing the EF from our analysis, we could not use many
other CVs, as there are not enough known inputs (as discussed, we need 5 input parameters to
perform our analysis). We have only accepted those inputs which have been
obtained independent of the EF.

\section{Constraining the modified parameter $ \Upsilon $}     
\label{comparison}

As mentioned earlier, the secondary masses are obtained in two ways. One from the observed data and the 
other via numerical analysis at different $ \Upsilon $ values. These two mass values are compared to find 
the range of the modified parameter for which the critical secondary masses best fit the observed values. In the 
numerical calculation, we use the polytropic index $ n=1.5 $ to model the secondary stars which move in circular
orbits in Schwarzschild backgrounds (the effects of rotation of the primary are negligible), and use a fourth
order approximation to the tidal potential. The choice of the polytropic index is
justified by the following facts. The secondary stars considered in this paper fall in the main sequence category. 
They also have small masses as reflected from their orbital periods which are less than $6$ hours. Such main 
sequence stars of masses below $ \sim 0.4 M_\odot $ are known to be highly convective and well described by a 
$ n=1.5 $ polytrope \cite{polytrope_1} \cite{polytrope_2}.
\begin{figure}
\centering
\includegraphics[scale=0.6]{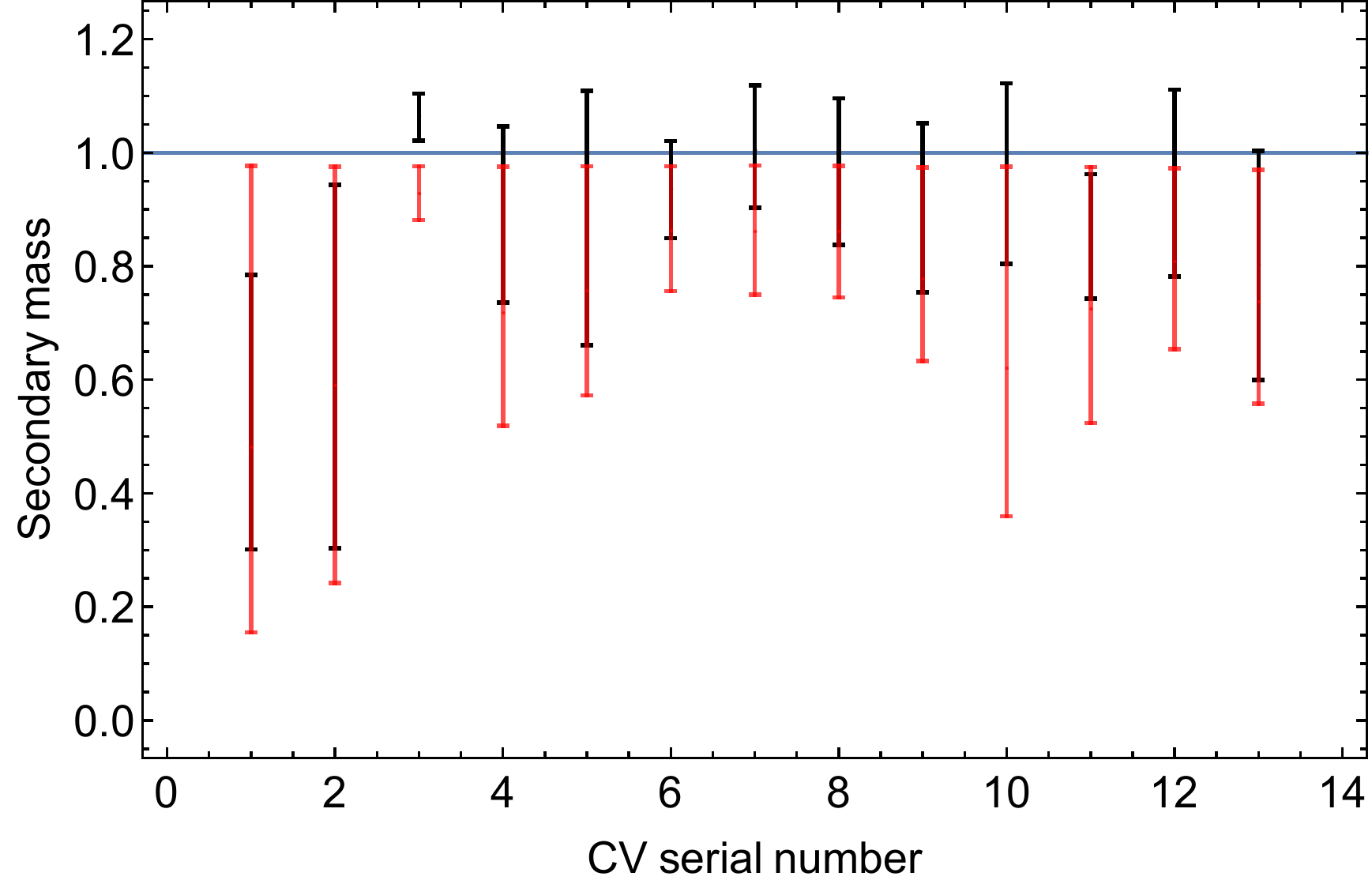}
\caption{Secondary masses for different CVs are plotted. Red bars indicate the critical mass range 
$ M_2^{crit} $ obtained for $ \Upsilon=0 $. Black bars indicate the mass range 
$ M_2 $ calculated from the observed data. The mass values of each system are divided by maximum values of $ M_2^{crit} $ 
at $ \Upsilon=0.47 $ which is suggested by the horizontal line. }
\label{fig2}
\end{figure}
Figure \ref{fig2} shows the two mass ranges together for all the CV systems. Black bars indicate the mass ranges 
obtained from Monte Carlo, whereas the red bars show the critical mass $ M_2^{crit} $ as obtained numerically at $ \Upsilon = 0 $. 
In order to do so, $ M_1 $, $ a $ and $ R_2 $, as generated by Monte
Carlo are used as inputs. These parameters are interrelated via Eqs. (\ref{MCE_1}) to (\ref{MCE_5}). 
As a result, the values of $ M_1 $, $ a $ and $ R_2 $, within their stipulated ranges (listed in
Table 2), can not be chosen independently of each other. We, therefore, use the sample values of 
$ M_1 $, $ a $ and $ R_2 $ as generated by Monte Carlo such that they fall within the $1\sigma $ limit.

To find the critical masses $ M_2^{crit} $ for all the sample sets of input parameters is numerically expensive. 
However, with the help of the EF, we can find the two sets of parameters that give the 
extreme values of the secondary mass $ M_2 $. These two sets of input parameters are then used in the 
numerical procedure to find the maximum and minimum critical masses $ M_2^{crit} $. The mean $ M_2^{crit} $ 
is calculated by using the mean values of $ M_1 $, $ a $ and $ R_2 $. It is fair to use the EF to find 
the two sets of input parameters since a set of parameters that gives the maximum secondary mass via the EF, also gives 
the maximum of all the $ M_2^{crit} $ values obtained via numerical calculations as well.
\begin{table}[h!]
\begin{center}
\small
\caption{Critical masses of the secondary stars at the tidal disruption limit (obtained via numerical analysis) 
for three different values of $ \Upsilon $. Here $ n=1.5 $. Max and min denote the maximum and minimum 
values of $ M_2^{crit} $ which are obtained using 
two different sets of values of the input parameters $ M_1 $, $ a $ and $ R_2 $. The mean value of $ M_2^{crit} $ is obtained 
using the mean of the input parameters. \label{table3}}
\begin{tabular}{cccccccc}
\hline
\hline
~~~~Name~~~~&~~~~$M_2$~~~~ & \multicolumn{3}{c}{ $ M_2^{crit}|_{\Upsilon=0} $ $ (M_\odot ) $ } & 
 $ M_2^{crit}|_{\Upsilon=0.47} $ $ (M_\odot ) $ & 
 $ M_2^{crit}|_{\Upsilon=0.54} $ $ (M_\odot ) $ \\
 & ~~~~~~~$(M_\odot)$~~~~~~~ & ~ min ~ & ~ mean ~ &  max  
& ~ max ~ &  ~max~   \\
\hline
V4140 Sgr & $0.10\pm0.05$ & $0.029$ & $0.090$ & $0.183$ & $0.188$ & $0.188$  \\
V2051 Oph & $0.22\pm0.11$ & $0.085$ & $0.206$ & $0.341$ & $0.350$ & $0.351$  \\ 
OY Car & $0.085\pm0.003$ & $0.071$ & $0.074$ & $0.078$ & $0.080$ & $0.080$  \\	
Ex Hya & $0.095\pm0.017$ & $0.056$ & $0.077$ & $0.104$ & $0.107$ & $0.107$  \\
HT Cas & $0.09\pm0.02$ & $0.060$ & $0.080$ & $0.103$ & $0.105$ & $0.106$  \\
IY Uma & $0.068\pm0.006$ & $0.055$ & $0.063$ & $0.071$ & $0.073$ & $0.073$  \\
Z Cha & $0.125\pm0.014$ & $0.093$ & $0.107$ & $0.122$ & $0.125$ & $0.125$  \\
DV Uma & $0.15\pm0.02$ & $0.116$ & $0.134$ & $0.152$ & $0.155$ & $0.156$  \\
IP Peg & $0.42\pm0.07$ & $0.297$ & $0.364$ & $0.457$ & $0.470$ & $0.471$ \\
UU Aqr & $0.44\pm0.07$ & $0.165$ & $0.284$ & $0.447$ & $0.458$ & $0.460$  \\
Gy Cnc & $0.36\pm0.05$ & $0.220$ & $0.304$ & $0.409$ & $0.420$ & $0.421$  \\
Ex Dra & $0.52\pm0.09$ & $0.357$ & $0.441$ & $0.530$ & $0.545$ & $0.548$  \\
V347 Pup & $0.53\pm0.13$ & $0.367$ & $0.484$ & $0.638$ & $0.657$ & $0.660$  \\
\hline
\end{tabular}
\end{center}
\end{table}
The critical masses resulted from numerical analysis at the tidal disruption limit are listed in Table \ref{table3} 
for $ 13 $ CV secondaries. From Fig. \ref{fig2}, it can be seen that, in general, the red bars cover the lower end 
of the black bars completely. On the other hand, for most of the systems, the upper ends of $ M_2^{crit} $ 
for $\Upsilon = 0$
remain less than the maximum observed values. As we know that the mass of the secondary increases with $\Upsilon$, 
the upper end of $ M_2^{crit} $ can be further increased by increasing $\Upsilon$. 
To be precise, we perform the numerical analysis with a 
non zero positive $ \Upsilon $ using the same set of $ M_1 $, $ a $ and $ R_2 $ which give the maximum value of 
$ M_2^{crit} $ at $ \Upsilon = 0 $. In this way, we ensure that the lowest possible value of the modified 
parameter is used.\footnote{This way we are able to isolate the effect of $\Upsilon$ and avoid possible degeneracy
with other parameters.} On the other hand, in view of the fact that the minimum limit of $ M_2^{crit} $ due to 
GR ($\Upsilon = 0$) already covers the lower part of the observed range, there is no need to decrease the value of the 
modified parameter. Thus we can set the lower limit of $ \Upsilon $ to zero to attain the minimum value of
the secondary mass calculated from the observed data. In order to put an upper 
limit on $ \Upsilon $, we perform the chi-square test. We define a quantity $ \chi^2 $ as 
\begin{equation}
\chi^2 =\sum_{i=1}^{N} \frac{\left(M_i^{obs}-M_i^{crit}|_\Upsilon \right)^2}{\sigma_i^2}~,
\end{equation}
where the subscript $ i $ stands for the $ i $-th CV system. Here, $ M_i^{obs} $ is the mean of the secondary 
mass $ M_2 $ obtained from observed data and $ M_i^{crit}|_\Upsilon $ is numerically calculated $ M_2^{crit} $ for 
a fixed $ \Upsilon $. It is to be mentioned again that $ M_i^{crit}|_\Upsilon $ is obtained at the set of input parameters 
$ M_1 $, $ a $ and $ R_2 $ which give the maximum value of $ M_i^{crit}|_{\Upsilon=0} $.
The uncertainty in $ M_i^{obs} $ is denoted by $ \sigma_i $. The total number of systems are denoted by 
$ N $, which in our case, is $ N=13 $. Now, to draw any conclusion, we need to find the degrees of 
freedom (d.o.f) in the chi-square analysis. Since the only parameter which is obtained from the data 
points of this test is $ \Upsilon $, d.o.f $= N-1 = 12$. Values of $ \Upsilon $ for which $ \chi^2$/d.o.f is less than 
one, are accepted. For $ \Upsilon=0 $ and $ 0.54 $, the values of $\chi^2$/d.o.f turn out to be $0.977$ and 
$1.017$ respectively. We find the upper limit at $ \Upsilon=0.47 $ for which $ \chi^2$/d.o.f is 
exactly $1$. In Fig. \ref{fig2}, all masses are shown in units of $ M_i^{crit}|_{\Upsilon=0.47} $, i.e., the maximum 
mass values corresponding to $ \Upsilon=0.47 $. There are other CV 
systems (e.g., BT Mon, WZ Sge, DQ Her, AE Aqr, etc.) for which binary parameters have been calculated. However, the 
secondary masses of these systems are decided to be outliers in our analysis. 

\section{Conclusions}
\label{conclusions}

One of the most important artifacts of modified gravity of the beyond-Horndeski class is that the
Vainshtein screening mechanism is only partially effective inside stellar objects, while it works
perfectly well outside these. This implies a modification of the pressure balance equation inside
stars characterized by a single parameter $\Upsilon$ in situations where there is no pressure anisotropy.
Since most stellar observables depend on such an equation, these theories 
therefore admit astrophysical tests. A considerable amount of literature has appeared in recent
years, that attempts to constrain $\Upsilon$ in these theories, using observational data on low mass stars. 

In this paper, we give a novel method to constrain $\Upsilon$ using tidal phenomena. We have focused
on cataclysmic variable binary stars, where a primary white dwarf accretes from a secondary donor. 
We have considered 13 such CV systems. Based on the available observational data of various binary parameters, 
we find the masses of the secondary stars in each of the CV systems. The secondary masses are then 
compared with that obtained from the numerical analysis for various values of the modified parameter. 
This comparison enables us to find a valid range of $\Upsilon$. In this paper, we find
$ 0 \le \Upsilon \le 0.47 $. Positive values of $\Upsilon$ are known to reduce the strength of 
gravity (compared to GR) inside stellar objects, which we find to be the case
inside donor stars in CV binaries. 
Here, we have used a fourth order approximation to the tidal potential. These fourth order effects become important,
as the ratio of the radius of the donor to that of the primary is not small. In this context, in our computations, the 
secondary donor star is assumed to move in a Schwarzschild background of the primary, and rotation effects of the 
primary are ignored. 
 
As an immediate application of the method discussed here, it would be interesting to understand 
how tidal effects in binary stars constrain Einstein-Born-Infeld gravity, which is a widely studied important modification of
GR. We leave such a study for the future.

\appendix
\section{Details on CV systems}  

\noindent
\textbf{V4140 Sgr}:
Borges and Baptista (2005) \cite{V4140_Sgr_1} give $ q $ and $ i $ using eclipse geometry. 
Mukai et al. (1988) \cite{V4140_Sgr_2} derive $ K_1 $ from emission line spectroscopy and 
Baptista et al. (1989) \cite{V4140_Sgr_3} determine $ \Delta\phi_{1/2} $.

\noindent
\textbf{V2051 Oph}:
Baptista et al. (1998) \cite{V2051_Oph_1} use the measurement of contact phases in eclipse light 
curve to obtain $ \Delta\phi_{1/2} $, $q $ and $ i $. Watts et al. (1986) \cite{V2051_Oph_2} give the 
value of $ K_1 $ from emission line which is in accordance with the velocity amplitude predicted by Baptista et al.

\noindent
\textbf{Oy Car}:
Wood et al. (1989) \cite{OY_Car_1} give mass ratio and the inclination angle using eclipse geometry. 
$ K_2 $ is measured by spectroscopic measurement by Copperwheat et al. (2012) \cite{Oy_Car_2}. 
This is consistent with photometric measurement of $ K_2 = 470 \pm 7 $ km~s$^{-1}$ by Littlefair et al. (2008) \cite{Oy_Car_3}. 

\noindent
\textbf{Ex Hya}:
Hellier et al. (1987) \cite{Ex_Hya_1} derive $ K_1 $ from radial velocity emission line. It is consistent with 
Gilliland's \cite{Ex_Hya_2} value of $ 58\pm9 $ km~s$^{-1}$ and also with $68\pm9$ km~s$^{-1}$ by Breysacher 
and Vogt (1980) \cite{Ex_Hya_3}. We prefer Hellier et al. as it has lower relative error and is the most recent 
work. $ K_2 $ was measured by Smith, Cameron and Tucknott (1993) \cite{Ex_Hya_4} using skew mapping 
technique. Vande Putte (2003) \cite{Ex_Hya_5} also gives $K_2 = 360\pm35 $ km~s$^{-1}$ which they 
conclude as not reliable. Hellier (1996) \cite{Ex_Hya_6} measures the inclination angle $ i $. The FWHM 
eclipse duration $ \Delta\phi_{1/2} $ is obtained by Ishida et al. (1994) \cite{Ex_Hya_7}. 

\noindent
\textbf{HT Cas}:
Horne et al. (1991) \cite{Ht_Cas_1} derive $ q $ and $ i $ from the observed eclipse duration 
$ \Delta\phi_{1/2} $ using eclipse geometry. They predict $ K_1 $ and $ K_2 $ from photometric 
model. This agrees with spectroscopic measurements of $ K_2 $.

\noindent
\textbf{IY Uma}:
Steeghs et al. (2003) \cite{IY_Uma_1} measure the mass ratio and inclination angle using eclipse 
geometry. This $ q $ value agrees with Patterson (2000) \cite{IY_Uma_2}. $ K_2 $ is obtained by 
Rolfe et al. (2002) \cite{IY_Uma_3} using skew mapping.

\noindent
\textbf{Z Cha}:
Wade et al. (1988) \cite{Z_Cha_1} derive $K_2$ from spectroscopic study. Mass ratio $ q $, 
$ \Delta\phi_{1/2} $ and inclination angle $ i $ is determined by Wood et al. (1986) \cite{Z_Cha_2} from 
eclipse geometry. Cook and Warner (1984) \cite{Z_Cha_3} find a larger mass ratio $ 5<1/q<7 $ for 
inclination angle $ 80.2^{\circ} < i < 82^{\circ} $. However, they do not consider the finite width of the 
stream. Therefore, it is not used.

\noindent
\textbf{DV Uma}:
Feline et al. (2004) \cite{DV_Uma_1} determine the system parameters by two methods. From the 
derivative of the light curve they obtain the eclipse contact phases of the white dwarf and the bright 
spot. The other one is a parametrized model of the eclipse fitted to the observed light curve by 
chi-square minimization. They prefer the second method over the derivative method. Using the 
parametrized method they find $ q $ and $ i $. Other than being consistent with Patterson et al. (2000) 
\cite{DV_Uma_2}, these values are more precise. The secondary is observed to fall in the spectral 
type M4.5$\pm$0.5 (Mukai et al. 1990 \cite{DV_Uma_3}). Accordingly, Henry et al. (1999) 
\cite{DV_Uma_4} give a value $0.12-0.17 M_\odot $. This is consistent with the value 
$ 0.15\pm0.02 M_\odot $ given by Patterson et al. (2000) via eclipse analysis.

\noindent
\textbf{IP Peg}:
Martin et al. (1989) \cite{IP_Peg_1} obtain $ K_2 $ from radial velocity curve by spectroscopic 
analysis. Smak (2002) \cite{IP_Peg_2} gives the mass ratio and the inclination angle from spot eclipse. 
However, the value of $ K_1 = 175\pm15$ km~s$^{-1}$ as determined from Marsh (1988) \cite{IP_Peg_3} 
gives a larger $ q $ value and therefore, not used. Using $ \Delta\phi_{1/2} $ from Wood and 
Crawford (1986) \cite{IP_Peg_4} and the other quantities we obtain the remaining system parameters. 

\noindent
\textbf{UU Aqr}:
Diaz and Steiner (1991) \cite{UU_Aqr_1} obtain the value $K_1$ from spectroscopic measurements. 
Vande Putte et al. (2003) \cite{Ex_Hya_5} determine $K_2$ from skew mapping. The values of $ K_1 $ 
and $ K_2 $ are compatible with inclination angle $ i $ and $ \Delta\Phi_{1/2} $ as given by 
Baptista et al. (1994) \cite{UU_Aqr_3} using eclipse geometry. 

\noindent
\textbf{Gy Cnc}:
Thorstensen (2000) \cite{Gy_Cnc_1} derives $ K_1 $ using emission line velocity amplitude. From 
spectroscopic analysis he obtains $ K_2=297\pm15 $ km~s$^{-1}$ which is corrected to $ 283\pm17$ km~s$^{-1}$. 
The mass ratio $ q $ is obtained using $ K_1 $ and $ K_2 $. Combining $ q $ with $ \Delta\phi_{1/2} $ he 
finds the inclination angle $ i $ from Roche lobe geometry. 

\noindent
\textbf{Ex Dra}:
Billington et al. (1996) \cite{Ex_Dra_1} obtain $ K_2 $ using H$\alpha $ emission line. They also 
measure the rotational broadening of secondary to measure $ v\sin i $. Baptista et al. (2000) \cite{Ex_Dra_2} 
obtain the mass ratio and inclination using stream trajectory from the secondary star to the position 
of the bright spot. Fiedler, Barwig and Mantel (1997) \cite{Ex_Dra_3} obtain $ K_1 $, $ K_2 $ and $ i $. 
However, systematic error in $ K_1 $ renders it unreliable.  

\noindent
\textbf{V347 Pup}:
Thoroughgood et al. (2005) \cite{V347_Pup_1} measure $ \Delta\phi_{1/2} $ by measuring the flux 
during eclipse. The value agrees with Buckley et al. (1990) \cite{V347_Pup_2}. They obtain $ K_2 $ 
using skew mapping with correction of the irradiation effect. They also calculate $ v\sin i $ using different 
spectral templates. Inclination angle is provided by Still et al. (1998) \cite{V347_Pup_3}.


\begin{thebibliography}{999}

\bibitem{CliftonRev}
T.~Clifton, P.~G.~Ferreira, A.~Padilla and C.~Skordis,
  Phys.\ Rept.\  {\bf 513}, 1 (2012).
\bibitem{LangloisRev}
D. Langlois, Int.\ J.\ Mod.\ Phys.\ D {\bf 28}, no. 05, 1942006 (2018).
\bibitem{IshakRev} 
M. Ishak, Living Rev.\ Rel.\  {\bf 22}, no. 1, 1 (2019).
\bibitem{KaseRev} 
R. Kase, S. Tsujikawa, Int.\ J.\ Mod.\ Phys.\ D {\bf 28}, no. 05, 1942005 (2018).
\bibitem{KobayashiRev}
T.~Kobayashi,
Rept.\ Prog.\ Phys.\  {\bf 82}, no. 8, 086901 (2019).
\bibitem{Nicolis}
A.~Nicolis, R.~Rattazzi and E.~Trincherini,
Phys.\ Rev.\ D {\bf 79}, 064036 (2009).
\bibitem{Deffayet} 
C.~Deffayet, G.~Esposito-Farese and A.~Vikman,
Phys.\ Rev.\ D {\bf 79}, 084003 (2009).
\bibitem{Deffayet1}
C.~Deffayet, X.~Gao, D.~A.~Steer and G.~Zahariade,
Phys.\ Rev.\ D {\bf 84}, 064039 (2011).
\bibitem{Kobayashi1}
T.~Kobayashi, M.~Yamaguchi and J.~Yokoyama,
Prog.\ Theor.\ Phys.\  {\bf 126}, 511 (2011).
\bibitem{Horndeski}
G. W. Horndeski, Int.\ J.\ Theor.\ Phys.\  {\bf 10}, 363 (1974).
\bibitem{BH1} 
J. Gleyzes, D. Langlois, F. Piazza, F. Vernizzi, Phys.\ Rev.\ Lett.\  {\bf 114}, no. 21, 211101 (2014).
\bibitem{BH2} 
J. Gleyzes, D. Langlois, F. Piazza, F. Vernizzi, JCAP {\bf 1502}, 018 (2015). 
\bibitem{LangloisNoui}
D.~Langlois and K.~Noui,
JCAP {\bf 1602}, 034 (2016).
\bibitem{LangloisNoui1}
D.~Langlois and K.~Noui,
JCAP {\bf 1607}, 016 (2016).
\bibitem{DHOST1}
M.~Zumalacárregui and J.~García-Bellido,
Phys.\ Rev.\ D {\bf 89}, 064046 (2014).
\bibitem{Vainshtein} 
A. I. Vainshtein, Phys. Lett. {\bf B 39} 393 (1972).
\bibitem{JainKhoury} 
B. Jain, J. Khoury, Annals Phys.\  {\bf 325}, 1479 (2010).
\bibitem{BabichevRev} 
E. Babichev, C. Deffayet, Class.\ Quant.\ Grav.\  {\bf 30}, 184001 (2013).
\bibitem{KWY} 
T. Kobayashi, Y. Watanabe, D. Yamauchi, Phys. Rev. {\bf D 91} 064013 (2015).
\bibitem{VainGW1}
M.~Crisostomi and K.~Koyama,
Phys.\ Rev.\ D {\bf 97}, no. 2, 021301 (2018)
\bibitem{VainGW2}
D.~Langlois, R.~Saito, D.~Yamauchi and K.~Noui,
  Phys.\ Rev.\ D {\bf 97}, no. 6, 061501 (2018).
\bibitem{SaksteinPRD} 
K. Koyama, J. Sakstein, Phys. Rev. {\bf D 91}, 124066 (2015).
\bibitem{Saito}
R.~Saito, D.~Yamauchi, S.~Mizuno, J.~Gleyzes and D.~Langlois,
  JCAP {\bf 1506}, 008 (2015).
\bibitem{Sakstein} 
J. Sakstein, Phys.\ Rev.\ Lett.\  {\bf 115}, 201101 (2015).
\bibitem{Sakstein3} Sakstein J., 2015b, Phys. Rev. D, 92, 124045
\bibitem{Jain} 
R. K. Jain, C. Kouvaris, N. G. Nielsen, Phys. Rev. Lett. {\bf 116}, 151103 (2016).
\bibitem{Saltas} 
I. D. Saltas, I. Sawicki, I. Lopes, JCAP {\bf 1805}, no. 05, 028 (2018).
\bibitem{Babichev}
E.~Babichev, K.~Koyama, D.~Langlois, R.~Saito and J.~Sakstein,
  Class.\ Quant.\ Grav.\  {\bf 33}, no. 23, 235014 (2016).
\bibitem{Tapo1}
S. Chowdhury, T. Sarkar, Ap. J {\bf 884}, 95 (2019).
\bibitem{GonzaloRev}
G.~J.~Olmo, D.~Rubiera-Garcia and A.~Wojnar,
arXiv:1912.05202 [gr-qc].
\bibitem{CVbook} B. Warner, {\tt Cataclysmic Variable Stars}, Cambridge Universty Press (1995).
\bibitem{Kopal} Z. Kopal, {\tt Close Binary Systems} (London: Chapman and Hall) (1953)
\bibitem{Pacynski} B. Pacynski, Annu. Rev. Astron. Astrophys., {\bf 9}, 183 (1971) 
\bibitem{Eggleton} P. P. Eggleton, Astrophys. J., {\bf 268}, 368 (1983)
\bibitem{Manasse-Misner} F. K. Manasse and C. W. Misner, J. Math. Phys., {\bf 4}, 735 (1963)
\bibitem{Ishii-Kerr} M. Ishii, M. Shibata and  Y. Mino, Phys. Rev. D, {\bf 71}, 044017 (2005)
\bibitem{BANERJEE201929} P. Banerjee, S. Paul, R. Shaikh and T. Sarkar, Phys. Lett. B, {\bf 795}, 29 (2019)
\bibitem{ecentricity} J. Patterson et al., Publ. Astron. Soc. Pac., {\bf 117}, 1204 (2005)
\bibitem{BT_Mon_1} D. A. Smith, V. S. Dhillon and T. R. Marsh, Mon. Not. R. Astron. Soc., {\bf 296}, 465 (1998)
\bibitem{V347_Pup_1} T. D. Thoroughgood, V. S. Dhillon, D. Steeghs et al., Mon. Not. R. Astron. Soc., {\bf 357}, 881 (2005)
\bibitem{Monte_Carlo_1} K. Horne, W. F. Welsh, R. A. Wade, Astrophys. J., {\bf 410}, 35 (1993)
\bibitem{R2_formula_1} K. Horne, H. H. Lanning, R. H. Gomer, Astrophys. J., {\bf 252}, 681 (1982)
\bibitem{R2_formula_2} W. R. Penning, D. H. Ferguson, J. T. McGraw, J. Liebert, R. F. Green, Astrophys. J., {\bf 276}, 233 (1984)
\bibitem{R2_formula_3} R. A. Downes, M. Mateo, P. Szkody, D. C. Jenner and B. Margon, Astrophys. J., {\bf 301}, 240 (1986)	
\bibitem{R2_formula_4} M. K. Harrop-Allin and B. Warner, Mon. Not. R. Astron. Soc., {\bf 279}, 219 (1996)
\bibitem{R2_formula_5} A. W. Shafter, Astron. J., {\bf 89}, 10 (1984)
\bibitem{R2_formula_6} V. S. Dhillon, T. R. Marsh and D. H. P. Jones, Mon. Not. R. Astron. Soc., {\bf 252}, 342 (1991)
\bibitem{R2_formula_7} S. B. Howell and S. A. Blanton, Astron. J., {\bf 106}, 1 (1993)
\bibitem{DV_Uma_1} W. J. Feline, V. S. Dhillon, T. R. Marsh and C. S. Brinkworth, Mon. Not. R. Astron. Soc., {\bf 355}, 1 (2004)
\bibitem{polytrope_1} S. Rappaport and P. C. Joss, Astrophys. J., {\bf 254}, 616 (1982)
\bibitem{polytrope_2} V. Renvoize, I. Baraffe, U. Kolb and H. Ritter, Astron. Astrophys., {\bf 389}, 485 (2002)
\bibitem{V4140_Sgr_1} B. W. Borges and R. Baptista, Astron. Astrophys., {\bf 437}, 235 (2005)
\bibitem{V4140_Sgr_2} K. Mukai, R. H. D. Corbet and A. P. Smale, Mon. Not. R. Astron. Soc., {\bf 234}, 291 (1988)
\bibitem{V4140_Sgr_3} R. Baptista, F. J. Jablonski and J. E. Steiner, Mon. Not. R. Astron. Soc., {\bf 241}, 631 (1989)	
\bibitem{V2051_Oph_1} R. Baptista, M. S. Catalan, K. Horne and D. Zilli, Mon. Not. R. Astron. Soc., {\bf 300}, 233 (1998)
\bibitem{V2051_Oph_2} D. J. Watts, J. Bailey, P. W. Hill et al., Astron. Astrophys., {\bf 154}, 197 (1986)
\bibitem{OY_Car_1} J. H. Wood, K. Horne, G. Berrinan and R. A. Wade, Astrophys. J., {\bf 341}, 974 (1989)
\bibitem{Oy_Car_2} C. N. Copperwheat, T. R. Marsh, S. G. Parsons et al., Mon. Not. R. Astron. Soc., {\bf 421}, 149 (2012)
\bibitem{Oy_Car_3} S. P. Littlefair, V. S. Dhillon, T. R. Marsh et al., Mon. Not. R. Astron. Soc., {\bf 388}, 1582 (2008) 
\bibitem{Ex_Hya_1} C. Hellier, K. O. Mason and S. R. Rosen, Mon. Not. R. Astron. Soc., {\bf 228}, 463 (1987)
\bibitem{Ex_Hya_2} R. L. Gilliland, Astrophys. J., {\bf 258}, 576 (1982)
\bibitem{Ex_Hya_3} J. Breysacher and N. Vogt, Astron. Astrophys., {\bf 87}, 349 (1980)
\bibitem{Ex_Hya_4} R. C. Smith, A.C. Cameron, D. S. Tucknott, in O. Regev, G. Shaviv eds, 
Cataclysmic Variables and Related Physics. IOP Publ., Bristol, p. 70 (1993)	
\bibitem{Ex_Hya_5} D. Vande Putte, R. C. Smith, N. A. Hawkins, J. S. Martin, Mon. Not. R. Astron. Soc., {\bf 342}, 151 (2003) 
\bibitem{Ex_Hya_6} C. Hellier, in A. Evans, J. H. Wood eds, Cataclysmic Variables and Related 
Objects. Kluwer Academic Publishers, Netherlands, p. 143 (1996)	
\bibitem{Ex_Hya_7} M. Ishida, K. Mukai and J. P. Osborne, Publ. Astron. Soc. Jap., {\bf 46}, L81 (1994)	
\bibitem{Ht_Cas_1}  K. Horne, J. H. Wood and R. F. Stiening, Astrophys. J., {\bf 378}, 271 (1991)
\bibitem{IY_Uma_1} D. Steeghs, M. A. Perryman, A. Reynolds, et al., Mon. Not. R. Astron. Soc., {\bf 339}, 810 (2003)
\bibitem{IY_Uma_2} J. Patterson, J. Kemp, L. Jensen, T. Vanmunster, D. R. Skillman, B. Martin, R.
Fried, J. R. Thorstensen, Publ. Astron. Soc. pac., {\bf 112}, 1567 (2000)
\bibitem{IY_Uma_3} D. J. Rolfe, T. M. C. Abbott and C. A. Haswell, Mon. Not. R. Astron. Soc., {\bf 334}, 699 (2002)						
\bibitem{Z_Cha_1} R. A. Wade and K. Horne, Astrophys. J., {\bf 324}, 411-430 (1988)	
\bibitem{Z_Cha_2} J. Wood, K. Horne, G. Berriman et al., Mon. Not. R. Astron. Soc., {\bf 219}, 629 (1986)
\bibitem{Z_Cha_3} M. C. Cook and B. Warner, Mon. Not. R. Astron. Soc., {\bf 207}, 705 (1984)
\bibitem{DV_Uma_2} J. Patterson, T. Vanmunster, D. R. Skillman, et al, Publ. Astron. Soc. Pac., {\bf 112}, 1584 (2000)	
\bibitem{DV_Uma_3} K. Mukai, et al., Mon. Not. R. Astron. Soc., {\bf 245}, 385 (1990)
\bibitem{DV_Uma_4} T. J. Henry, et al., Astrophys. J., {\bf 512}, 864 (1999)
\bibitem{IP_Peg_1} J. S. Martin, M. T. Friend, R. C. Smith and D. H. P. Jones, Mon. Not. R. Astron. Soc., {\bf 240}, 519 (1989)
\bibitem{IP_Peg_2} J. I. Smak, Acta Astronomica, {\bf 52}, 189 (2002)
\bibitem{IP_Peg_3} T. R. Marsh, Mon. Not. R. Astron. Soc., {\bf 231}, 1117 (1988)
\bibitem{IP_Peg_4} J. Wood and C. S. Crawford, Mon. Not. R. Astron. Soc., {\bf 222}, 645 (1986)
\bibitem{UU_Aqr_1} N. P. Diaz and J. E. Steiner, Astron. J., {\bf 102}, 1417 (1991)
\bibitem{UU_Aqr_3} R. Baptista, J. E. Steiner and D. Ciesninski, Astrophys. J., {\bf 433}, 332 (1994)
\bibitem{Gy_Cnc_1} J. R. Thorstensen, Publ. Astron. Soc. pac., {\bf 112}, 1269 (2000)
\bibitem{Ex_Dra_1} I. Billington, T. R. Marsh and V. S. Dhillon, Mon. Not. R. Astron. Soc., {\bf 278}, 673 (1996)
\bibitem{Ex_Dra_2} R. Baptista, M. S. Catalan and  L. Costa, Mon. Not. R. Astron. Soc., {\bf 316}, 529 (2000)
\bibitem{Ex_Dra_3}  H. Fiedler, H. Barwig and K. H. Mantel, Astron. Astrophys., {\bf 327}, 173 (1997)
\bibitem{V347_Pup_2} D. A. H. Buckley, D. J. Sullivan, R. A. Remillard, I. R. Tuohy, M. Clark, Astrophys. J., {\bf 355}, 617 (1990)
\bibitem{V347_Pup_3} M. D. Still, D. A. H. Buckle and M. A. Garlick, Mon. Not. R. Astron. Soc., {\bf 299}, 545 (1998)
		
\end{thebibliography}
\end{document}